# オープン IoT に向けたヘテロリソースの最適化検討

山登庸次[†]　　干川尚人[†]　　野口博史[†]　　出水達也[†]　　片岡操[†]

† NTT ネットワークサービスシステム研究所
東京都武蔵野市緑町 3-9-11
E-mail: †yamato.yoji@lab.ntt.co.jp

**あらまし**　近年，IoT 技術が進展しており，数多くのセンサ，アクチュエータがネットワークに繋がってきている．従来，IoT サービスは，垂直統合的に開発がされていたが，より多彩なサービスを実現するためには，デバイスとサービスを水平分離的に相互に利活用する，オープン IoT が重視されてきている．私達は，オープン IoT に向け，ユーザが必要なデータを持つデバイスをオンデマンドに発見し，利用する，Tacit Computing 技術を提案し，その要素技術を実装してきた．本稿では，Tacit Computing で発見した，ユーザに必要なデータを持つデバイスを継続的に利用するサービスとするため，運用コスト低減や性能向上を行うための 3 レイヤーの最適化を提案する．最適化では，デバイス，ネットワーク，クラウドレイヤーで，適切な機能配置等を，本格運用前に行う．
**キーワード**　IoT, Tacit Computing，オープン IoT，マルチレイヤー，最適化，ユーザコンテキスト

## A Study of Optimizing Heterogeneous Resources for Open IoT

Yoji YAMATO[†], Naoto HOSHIKAWA[†], Hirofumi NOGUCHI[†], Tatsuya DEMIZU[†], and Misao KATAOKA[†]

† Network Service Systems Laboratories, NTT Corporation
3-9-11, Midori-cho, Musashino-shi, Tokyo 1808585 Japan
E-mail: †yamato.yoji@lab.ntt.co.jp

**Abstract**　Recently, IoT technologies have been progressed, and many sensors and actuators are connected to networks. Previously, IoT services were developed by vertical integration style. But now Open IoT concept has attracted attentions which achieves various IoT services by integrating horizontal separated devices and services. For Open IoT era, we have proposed the Tacit Computing technology to discover the devices with necessary data for users on demand and use them dynamically. We also implemented elemental technologies of Tacit Computing. In this paper, we propose three layers optimizations to reduce operation cost and improve performance of Tacit computing service, in order to make as a continuous service of discovered devices by Tacit Computing. In optimization process, appropriate function allocations are calculated for device, network and cloud layer before full-scale operation.
**Key words**　IoT, Tacit Computing, Open IoT, Multi Layer, Optimization, User Context

## 1. はじめに

近年，IoT（Internet of Things）技術が発展しており（例えば，[1]-[8]），数多くのセンサやアクチュエータ，ロボット等の IoT 機器がネットワークに繋がってきている．IoT とは，物体に通信機能を持たせ，ネットワークに接続し，データを処理分析することで，自動制御等を行うための技術である．IoT の適用範囲は多岐に渡り，Industrie4.0 [9] で検討されている製造，流通，メンテナンスだけでなく，健康管理，医療，農業，エネルギー等も適用先として期待されている．

しかし，現状，IoT の適用は，特定のターゲットに合わせた垂直統合的なワンオフソリューションになりがちである．センサの選定，分析，表示，アクションを一括で行い，システム化して提供する SI ビジネスが多い．この結果，個々の IoT サービスのコストが高くなり，多彩なサービスが十分出てきていない現状がある．

多彩な IoT サービスをコスト低く開発，運用するためには，センサやアクチュエータ等のデバイスとサービスを分離し，水

平分離的にデバイスとサービスを相互に利用可能にすることが必要となる．このような考えはオープン IoT と呼ばれ，今後の IoT サービス活性化に重視されている．

私達は，このようなオープン IoT に向けて，サービスからデバイスを自由に利用するための仕組みとして，Tacit Computing 技術を提案している [10]．Tacit Computing は，デバイスが今持つライブデータに基づき，ユーザが必要なデータを持つデバイスをオンデマンドに発見し，利用することを可能にする技術であり，私たちはその要素技術の一部を実装してきている．

しかし，その時点の状況に合わせてデバイスを発見し，利用するだけでは，ユーザのニーズには一時しか答えているとは言えない．ユーザのニーズに応え続けるサービスとするためには，ユーザに必要なデータを持つデバイスを継続的に利用することをリーズナブルな価格で提供することが必要である．そこで，本稿では，Tacit Computing で発見利用するデバイスを継続的に利用したサービスとするため，運用コスト低減や性能向上を行う 3 レイヤーの最適化を提案する．最適化では，デバイス，ネットワーク，クラウドレイヤーで，適切な機能配置等を，サービスの本格運用前に行う．本稿は，3 レイヤーの最適化のアプローチを述べ，個々の最適化要素技術の詳細評価は別稿で行う．

本稿の構成は以下の通りである．2 節で，私たちが提案している Tacit Computing の概要を説明する．3 節で，発見したデバイスを継続的に利用するサービス化のための課題を挙げる．4 節で，3 レイヤーで最適化するためのアプローチを述べる．5 節で本稿をまとめる．

## 2. Tacit Computing の概要

本節では，私達が提案している Tacit Computing の概要を図 1 を用いて説明する．

Tacit Computing では，クラウドレイヤー，ネットワークレイヤー，デバイスレイヤーの 3 層から，ユーザに適切なリソースを発見，連携することで，ユーザのリクエストに応える．3 層の構成であるが，時々刻々変化する状況に対応するため，現場に近いデバイスレイヤーでできるだけ処理を行うのがコンセプトである．低レイヤーで処理を終えることで，ネットワークトラフィックの削減や，プライバシー性が高いデータの流出を抑えることができる．

Tacit Computing のコンセプトを提案し，キーとなる要素技術を実装し，NTT R&D フォーラム 2017 [11] に展示した．本節では，特にオープン IoT 時代のキーとなる，ライブデータ検索技術とデバイス仮想化技術について述べる．

ライブデータ検索技術は，ユーザにとって必要なデータを提供するデバイスを検索するための技術である．IoT サービスの例として，橋等の施設にセンサを複数設置し，劣化状況等をモニタリングするといったサービスがある．この場合は，劣化状況がすぐ進むとは考えられないため，センサデータは，数時間等の周期で複数点の状況をクラウドにデータを上げ，劣化状況の変化を統計ソフトや機械学習で分析すれば良い．一方，定点のカメラに映った人物に対して，情報案内や警告アラート等を

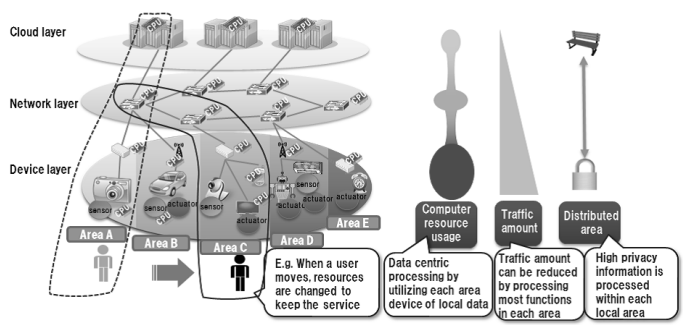

図 1　Outline of Tacit Computing

行うような例では，数秒程度しかカメラに人物は映らないし，また，そのカメラの映像しかその人物には意味がないデータと言える．このように，デバイスレイヤーで発生し，時々刻々変わるようなデータをライブデータと呼ぶ．

そこで，ユーザにとって必要なライブデータを検索するため，クラウドレイヤーにデータが上がって来るのを待つのではなく，下位レイヤーに分析する機能を配信することを，ライブデータ検索技術では行う．

例えば，箱根駅伝予選会に友人が出ており，友人が映ったカメラの映像を自動で繋いで欲しいとする．この場合，友人のゼッケン番号を検索キーに，リクエストをすると，Tacit Computing では，カメラを収容するゲートウェイやネットワークエッジに，OpenCV [12] 等の画像を分析する機能を配信して，カメラに近い場所で映像を分析することで，友人のゼッケン番号が画像分析で抽出され，友人が映っているカメラを特定する．

次に，利用したいデバイスが特定された場合に，そのデバイスを利用する必要がある．IoT デバイスは，多数のメーカーが開発しており，利用時のプロトコルやインタフェース，アドレス等が，デバイス毎に異なる．そこで，デバイス仮想化技術によって，個々のデバイスのインタフェース等の違いを吸収している．例えば，上記例であれば，利用方法はカメラ毎に異なるが，Tacit Computing では，カメラ映像の取得のような共通的なリクエストを元に，カメラを収容するゲートウェイ等で，デバイス毎のアダプタにて変換を行い，個々のカメラに応じたリクエストを行う．このようにすることで，ユーザはデバイス個々の違いを意識せず，デバイスの利用が可能である．また，デバイス利用時は Semantic を用いて抽象化する Semantic Web Services 等の既存技術も利用できる（例えば，[13]- [34]）．

## 3. 本稿で取り組む課題

2 節で述べた，Tacit Computng により，分析プログラムの配信等で，ユーザが必要なライブデータを持つデバイスを発見し，デバイス仮想化の仕組みを用いることで，個々のデバイスのインタフェース等が隠蔽されるため，ユーザはデバイスのアドホックな利用が可能となる．（勿論，オープン IoT の時代では，デバイスの発見利用の前に，デバイスの他者利用の社会的合意や，利用に伴う認証認可課金等の仕組みが必要となる．）

しかし，Tacit Computing を使って，その時点の状況に合わ

せてデバイスを発見し，利用するだけでは，ユーザのニーズには一時しか答えていないと考える．

Tacit Computing の利用例として，追跡カメラを考える．追跡カメラは，小学生等の子供の，学内や登下校中の映像を，子供近傍のカメラから取得し，親の携帯端末で見る利用形態である．追跡カメラで，親が子供の安全を映像で確認することはその時点の親のニーズは満足できるが，それを 1 回幾らという課金とした場合に受け入れづらいと考える．むしろ，子供近傍のカメラから映像を取得し続けるが，親は携帯端末で見たい時は子供の映像を見られ，見ていない時にも機械学習等で見守り，登下校中に不審者が近づいた等の異常時は親にアラートを送る等の，継続的なサービスとすることで，月額幾らと言った課金とした場合にも受け入れられると考える．

追跡カメラの処理としては，Tacit Computing で，子供が映っているカメラを発見し，親のリクエスト時はそのカメラの映像を親の携帯端末に届けるとともに，見守りのため，カメラを収容するゲートウェイまたはネットワークのエッジの SSE（Subscriber Service Edge）等に画像処理をする機能（OpenCV ライブラリ等）を配置し，子供が映っているカメラの映像から画像を切り出し分析を行う．画像分析した結果は，特徴ベクトルにサマライズされた後集約され，クラウド技術を用いて（例えば，[35]- [54]），不審者が近づいている等の異常値分析を機械学習等の手法（Local Outlier Factor 等 [55]）を用いて行い，異常がある場合は親にアラートする．

追跡カメラ自体は，Tacit Computing でなくても，カメラと機械学習等を使った SI サービスとしても作りこむことができるため，「Tacit Computing で発見利用するデバイスを継続的に利用したサービスをリーズナブルに提供すること」を，今後の課題と設定する．

Tacit Computing で，最初に，デバイスをアドホックに発見，利用する際は，まず使えることが必要であるため，コストや性能は度外視されている．しかし，それを継続的なサービスとしてリーズナブルに提供するためには，運用コスト低減や性能向上が必要となってくる．

そこで，本稿では，そのために，デバイス，ネットワーク，クラウドのレイヤーで，機能配置を適切に行うことによる，最適化を提案する．最適化は，最初のアドホックなデバイス利用期間中に行うことを想定している．例えば，追跡カメラの場合では，初日は試し利用などの形で子供近傍のカメラ映像を携帯端末で確認をできるようにし，その間にネットワークやクラウドでの最適な機能配置を計算し，設定することで，翌日以降は，リーズナブルな価格で追跡カメラサービスを提供できるようにする．

## 4. 3 レイヤーでの最適化提案

本節では，Tacit Computing によるデバイス利用を継続的なサービスとする際の，コスト削減，性能向上のための最適化のアプローチを述べる．最適化は，デバイス，ネットワーク，クラウドの各レイヤーでの処理と共に，レイヤー間で機能配置を変更するレイヤーを跨ぐ処理がある（図 2）．

デバイスレイヤーでは，まず，ユーザの要望に合うデバイスへの切換を随時行っていくことが必要である．追跡カメラであれば，子供の場所に応じて子供の映像が映っているカメラを選択していくことが必要である．この時，ネットワークに繋がる全カメラで画像解析をし，子供が映っているカメラだけ使うのでは，非常にコストが高くなってしまうため，子供を何らかの識別子を振り，カメラのメタデータを用いて，その識別子の子供の近傍のカメラを絞り込んでいく処理が必要となる．IoT デバイスは非 IP の機器も多いため，非 IP の機器を含めて，IoT デバイスの通信パターンを解析し，メタデータを自動か半自動で付与していくことを検討している．

更に，デバイスレイヤーでは，デバイスを収容するゲートウェイ群の中からどのゲートウェイに分析機能を配置するかが，コストに大きく影響する．追跡カメラであれば，OpenCV 等の画像分析機能が使われるが，よく子供が通るエリアのゲートウェイには画像分析機能を配置し，そのエリアを超えて移動する場合は，画像分析機能を配信して処理する等を検討している．

ネットワークレイヤーでは，ゲートウェイでの機能配置と同様に，SSE 等のネットワークエッジへの機能配置について最適化を行う事を検討している．ネットワークでの情報のキャッシュや配置については，情報セントリックネットワーク [56] 等で検討がされており，それらの適用も含めて検討する．また，サービスによっては，ネットワークで，帯域確保や優先制御等が必要となる場合もある．例えば，NGN（Next Generation Network）であれば，RACF（Resource and Admission Control Function）という帯域確保のための機能があり，そのようなネットワーク側の設定準備も，最適化処理の期間中に行う必要がある．

クラウドレイヤーでは，クラウドのどこで処理を行うかが，コストや性能に大きく影響する．利用頻度が多いデバイスを収容するネットワークエッジとの遅延が小さい DC（Data Center）のクラウドに処理機能をデプロイする．更に，クラウドリソースのサイズも運用コストに影響があるため，そのサイズも最適化対象である．例えば，追跡カメラであれば，サマライズされた特徴ベクトルを，機械学習などの手法で分析し，不審者のアラート等を行うことがクラウドでの処理で考えられるが，Storm [57] や Spark [58] のようなストリーム処理機能で処理するための適切なリソースサイズを確保する．

更に，ここまで，デバイスレイヤー，ネットワークレイヤー，クラウドレイヤー個々について述べていたが，サービスの利用形態によって，どこに機能を配置するのが適切かは変わってくる．そこで，各処理を，各レイヤーで処理する場合の，コスト（リソース確保のためのサーバコスト，発生トラフィックに基づくネットワークコスト）と性能（スループット，遅延）をモデル化し，どこに配置するのが適切かを計算することを検討している．ユーザ向けの価格・コストを制約条件として性能の最適化をするため，組合せ最適化問題になるため，限られた最適化期間中にある程度の解は発見できるような短時間での手法で計算を行う．

本節で述べた，各レイヤーの最適化，及び，3 レイヤーの配置

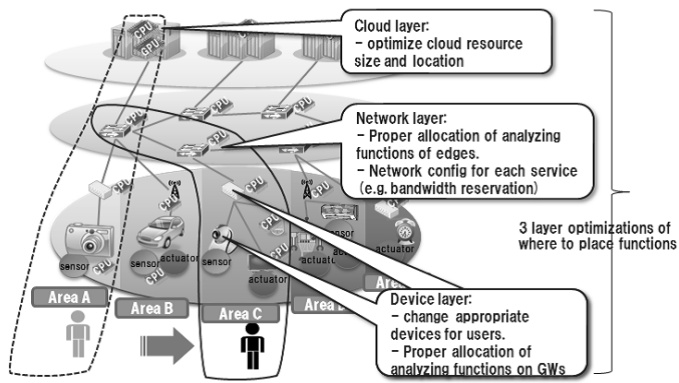

図 2　Three layer optimizations

最適化について，詳細評価を今後行い，別稿で報告予定である．

## 5. ま と め

本稿では，オープン IoT に向け，ユーザが必要なデータを持つデバイスをオンデマンドに発見し，利用する，Tacit Computing 技術の次のステップとして，3 レイヤーでの最適化処理を提案した．Tacit Computing で発見した，ユーザに必要なデータを持つデバイスを継続的に利用するサービスとするため，運用コスト低減や性能向上によるリーズナブルな提供が必要であり，デバイス，ネットワーク，クラウドレイヤーで，適切な機能配置等を，本格運用前に行う．

追跡カメラを一例に 3 レイヤーでの最適化アプローチについて検討を行った．デバイスレイヤーでは，デバイスの位置等のメタデータを自動把握し，適切なカメラに切り換えていくことが必要である．デバイスレイヤーのゲートウェイやネットワークレイヤーのエッジでは，分析処理する機能の配置によって，ネットワークコストや遅延が大きく影響するため，利用される可能性が高いカメラを収容するゲートウェイやエッジへは分析機能を配置し，可能性が低い場所には配置しない等の，計算を行う．クラウドレイヤーでは，通信遅延が低い DC のサーバへの機能配置等を行う．

今後は，本稿で提案した 3 レイヤーの最適化について，詳細評価を行い，別稿で報告する．また，Tacit Computing と最適化を用いて，実デバイスを管理，利用する実験を進める予定である．